\documentclass{aa}
\usepackage{graphicx}
\begin{document}
\title{Variable stars in nearby galaxies.\thanks{Based on
observations collected at ESO-La Silla}}
\subtitle{VI. Frequency-period distribution of Cepheids 
in IC 1613 and other galaxies of the Local Group}
\author{E. Antonello \and D. Fugazza \and L. Mantegazza}
\institute{Osservatorio Astronomico di Brera, Via E.~Bianchi 46,
       I--23807 Merate, Italy}
\offprints{E. Antonello} 
\date{Received 29 January 2002; accepted 29 March 2002}
\titlerunning{Cepheid frequency--period distribution}
\authorrunning{E. Antonello et al.}

\abstract
{The frequency--period distribution and other properties of Cepheids in 
IC 1613 are discussed and compared with those of stars in our Galaxy (Milky
Way), LMC, SMC, M31 and M33. Taking into account the observational limitations
and related incompleteness, it is concluded that the frequency--period 
distribution of Cepheids in IC 1613 is similar to that of SMC; we suspect 
that a much larger number of stars exist in IC 1613 with a period of less 
than 2 d that have not yet been detected. A discussion of the deficiency of 
fundamental mode Cepheids with periods in the range 8 - 10 d in the Milky Way,
M31 and M33 is reported. 
The present data are not sufficient to verify if this is produced by  
a real bimodal frequency--period distribution or whether depends on the lack 
of pulsating stars in such a period range due to pulsational stability 
reasons. Some arguments are presented in favor of a bimodal distribution
that is a function of the average metallicity. The Milky Way, M31 and M33 
have the two maxima located at the same periods, about 5 and 13 d, 
respectively. A comment on very long period Cepheids is also given.
}

\maketitle

\section{Introduction}
Cepheids are variable stars that are used to measure distances of galaxies
in the Local Group and nearby clusters, and are the primary calibrators for
the secondary standard candles that are applied at much greater distances.
However, they are important also for testing the theories on the internal
constitution of stars and stellar evolution. In particular, the study of
Cepheids in nearby galaxies is important to understand
the effects of different metallicities and corresponding
mass--luminosity relations on the pulsational characteristics. 
The purpose of our project was to obtain good light curves of Cepheids in 
IC 1613 and NGC 6822. In order to make best use of telescope time and 
reach fainter magnitudes, our strategy was to observe in white light,
i.e. without filter ($Wh$ photometry). The results of the survey for
variable stars in four fields of IC 1613 were reported in the previous
papers of this series (see Mantegazza et al. \cite{pa4}, Paper IV,
and references therein), while the properties of population I and II 
Cepheids observed in a field were discussed by Antonello et al. 
(\cite{pa2}, Paper II). Recently, the first results for NGC 6822 have
also been presented (Antonello et al. \cite{pa5}, Paper V). In another paper 
(Antonello et al. \cite{ant1}), the light curves of 
the Cepheids with period near 10 d of IC 1613 were analyzed 
and compared with those of the corresponding objects in the Galaxy and in 
the Magellanic Clouds. In the present paper we summarize the characteristics 
of the Cepheids observed in IC1613 and make a comparison with other galaxies 
containing a large number of such stars, taking
into account the results of other recent surveys. We will update the 
discussion about observational evidence for a dip in the frequency--period 
distribution near 8 - 10 d, that is seen preferentially in metal rich 
galaxies. Presently there are three different interpretations of such dips. 

1) In their extensive study on the frequency-period distribution of Cepheid 
variables, Becker et al. (\cite{bit}; hereafter BIT) concluded that it is 
not compatible with a standard birthrate function, and can only be explained 
if an ad hoc two-component birthrate function is adopted. The primary
component is a time--averaged ``background'' rate which reproduces the main
peak of the distribution (e.g. that in the Milky Way). The second component is 
produced by recent star formation (e.g. in OB associations) and is 
characterized by some cutoff below a critical mass; this component 
reproduces the secondary peak and the cutoff explains the dip.

2) Chiosi (\cite{chi}) noted that, since the blue band of core He-burning 
of models with relatively large mass ($\ga 9 M_{\sun}$) tend to move back 
to the red giant region (a tendency enhanced by increasing metallicity), 
such masses would spend a large fraction of the blue band lifetime within 
the instability strip; therefore the excess of long $P$ Cepheids would 
be intrinsic to the stellar models and a two-component birthrate is not 
needed. 

3) Buchler et al. (\cite{bu1}) suggested that the dip could be a real 
deficiency of stars caused by the instability of the nonlinear fundamental 
pulsation cycle in this period range, and such stars that cannot pulsate 
in the fundamental mode actually pulsate in the first overtone one. 

We recall a wish expressed more than twenty years ago by BIT (p. 649) 
in relation to the Cepheids in the seven galaxies, Milky Way, LMC, SMC, M31, 
M33, IC 1613 and NGC 6822: ``The Cepheids in M33, NGC 6822 and IC 1613 have 
only begun to be studied, and much more work needs to be done. Further 
information on the frequency--period distribution of any of these seven 
galaxies would be very helpful''.

\begin{figure}
 \includegraphics[width=15cm]{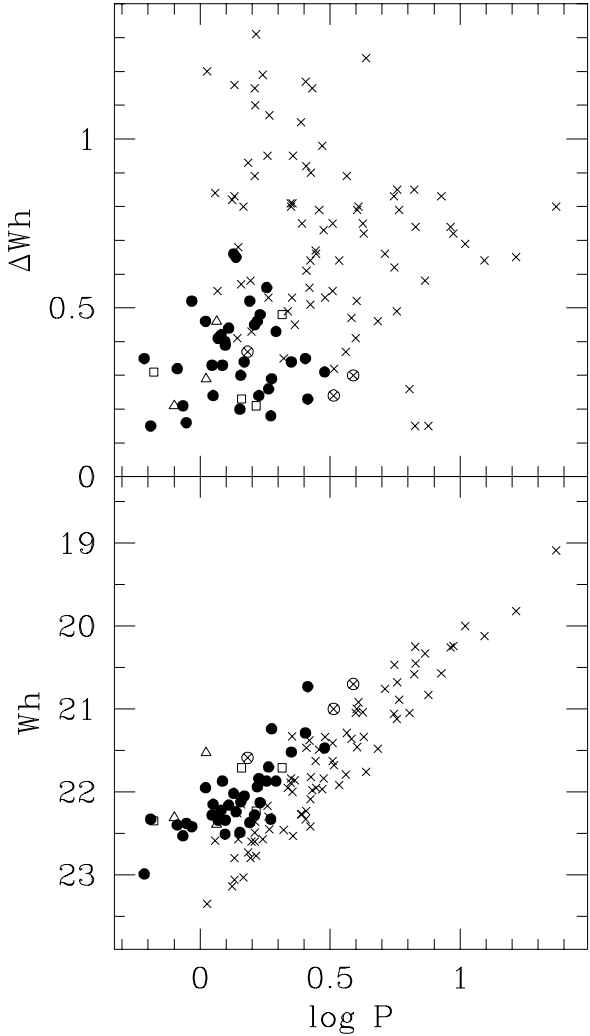}
 \caption[ ]{{\em Lower panel}: $PL$ relation for IC 1613 Cepheids. 
 {\em Crosses}: fundamental
 mode; {\em filled circles}: first overtone mode; {\em open squares}: uncertain
 mode; {\em open triangle}: second overtone candidate; {\em cross in circle}:
 fundamental mode Cepheid with a probable companion. For a comparison  
 with other surveys, we remark that $V \sim Wh+0.3$ for stars with cepheid
 color. {\em Upper panel}: full amplitude against $P$ }
 \label{apl}
\end{figure}

\section{Cepheids in IC 1613}

\subsection{Generalities}
Details of the $Wh$ CCD observations (performed with the ESO Dutch 0.9 m 
telescope) and reduction methods were reported in the previous papers of 
this series, along with the lists of variable stars and their coordinatae 
(Antonello et al \cite{pa1}; Antonello et al. \cite{pa3}; Mantegazza et al. 
\cite{pa4}). The finding charts of the variable stars are available upon 
request to the authors, or can be found in the Archive at the site
http://www.merate.mi.astro.it/$\sim$fugazza/cepheids.html.
The analysis of the $Wh$ light curves was performed as in Paper II, and
the Fourier parameters were used to identify the pulsation mode of
shorter $P$ Cepheids. The mode classification of the objects for each field 
was reported 
in the previous papers; we have however revised it by using homogeneous
criteria, and in particular we tried to classify the stars previously 
reported with uncertain modes. The result is that the $F$ mode pulsators
include also the following stars: V0236A, V2991B, V3277B, V3782B, V0793D, 
V1307D, and the $1OT$ pulsators include also the following stars: V0178A, 
V0524A, V1296A, V1289A, V0223B, V0987B. We recall that, owing to the 
uncertainties on the parameters, the mode identification
for the shorter $P$ stars is only preliminary.

In the lower panel of Fig. \ref{apl} the $PL$ relation is shown for the 
whole set of stars, which includes 82 $F$ mode, 34 $1OT$ mode,
3 suspected second overtone mode pulsators, and 5 Cepheids with uncertain mode.
In the upper panel we show the diagram of amplitude vs. $P$. 
The three $F$ mode Cepheids marked with a cross in a circle have the lowest 
amplitude and are the brightest stars among those with similar $P$; we suspect 
that they are binary systems or there is a blending effect.
An analysis of all the Fourier parameters indicates that, for $P$ less than 
about 4 d, the formal errors can be relatively large, and the resulting 
relation between the phase difference $\phi_{21}=\phi_2 - 2\phi_1$ (of the 
second and first Fourier component) and $P$ is relatively scattered. 
Therefore, this does not allow us to compare reliably such
parameters of short $P$ fundamental mode Cepheids in IC 1613 with those of 
similar stars in other galaxies, when looking for slight differences due to 
different metallicities. In other words, with the present data, useful 
comparisons should be limited to stars with $P \ga 4$ d (e.g.
Antonello et al. \cite{ant1}), instead of attempting a general comparison,
as done in Paper II.

\subsection{Comparison with OGLE survey}
Recently, the OGLE project made a survey of Cepheids in IC 1613, in the 
$V$ and 
$I$ bands (Udalski et al. \cite{uda5}), using a 1.3 m telescope. Owing to the 
larger field of view (14$\farcm$), they observed almost the whole galaxy and 
found about 135 population I Cepheids, with a limiting average magnitude 
$V \sim 22.6$. In the $V$ band the total exposure time was 900 sec, and 
the seeing conditions were much better than those of our project; in fact 
the Dutch 0.9 m telescope suffered from focusing 
problems. In our four selected fields we found about 128 Cepheids, with a
limiting magnitude of $V \sim 23$. We verified that, in the same four 
fields, OGLE detected about 75 Cepheids; there are 64 Cepheids in common 
with us. The comparison shows that there are systematic differences of some 
arcsec in the declination and right ascension between OGLE and our lists.
The $P$ values are reasonably in agreement. The largest discrepancies 
were found for three stars with $P$ close to 2 d, in which case it is 
difficult to distinguish between the true $P$ and the alias produced by the 
one--day separation of observations. Indeed, in the power spectrum there 
is a significant peak at both $f=1/P$ d$^{-1}$ and
$f=1-1/P$ d$^{-1}$. We merged our and OGLE dataset after 
correcting for the different mean magnitude and different amplitude, 
and we got for $V0122B$=$OGLE12044$ 2.032 d, for $V0987B$=$OGLE05209$ 2.094 d, 
and for $V0231C$=$OGLE01092$ 1.802 d. 

We also analysed several merged data sets, containing about 90 -- 100 $Wh$
and $V$-corrected data points each, but we were not able to improve the
accuracy of the light curve Fourier parameters of shorter $P$ stars. In 
particular we cannot confirm the possible indications for the resonance 
effects in $1OT$ pulsators reported in Paper II. 
The conclusion is that the accurate 
study of the light curves of Cepheids fainter than $V \sim 22$ requires a 
larger telescope than that adopted by us or by the OGLE project.

Finally we mention the deep HST imaging of a field in the halo of IC 1613 
performed by Dolphin et al. (\cite{dol}); they found 11 short $P$ Cepheids and
13 RR Lyrae stars. 

\section{Period distribution}
\begin{figure}
\resizebox{\vsize}{!}{\includegraphics{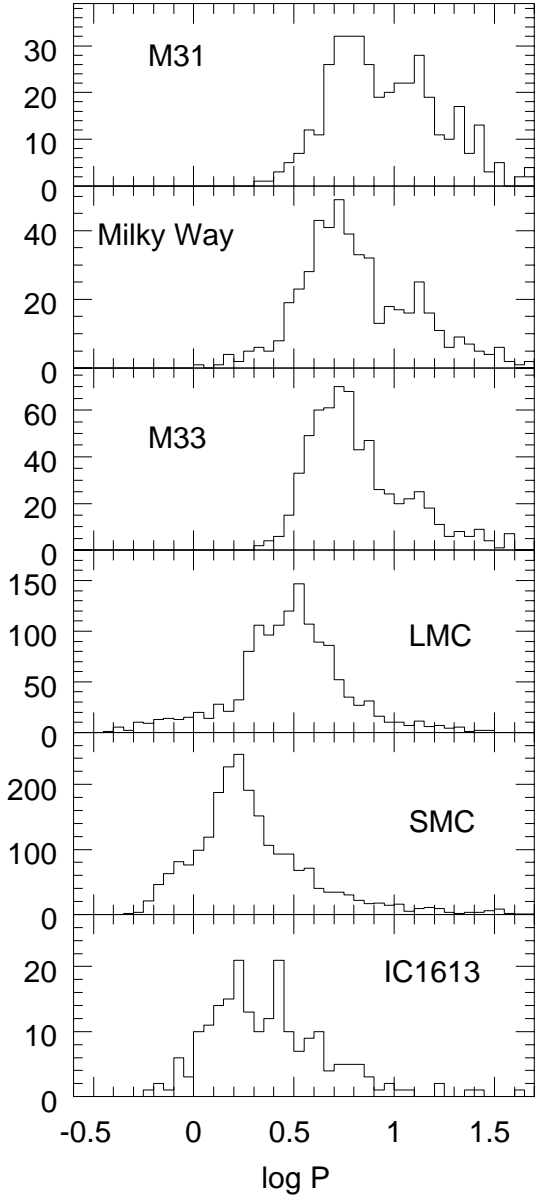}}
 \caption[ ]{Period distribution of Cepheids in galaxies of the 
Local Group
}
 \label{isto}
\end{figure}

\begin{figure}
\resizebox{\vsize}{!}{\includegraphics{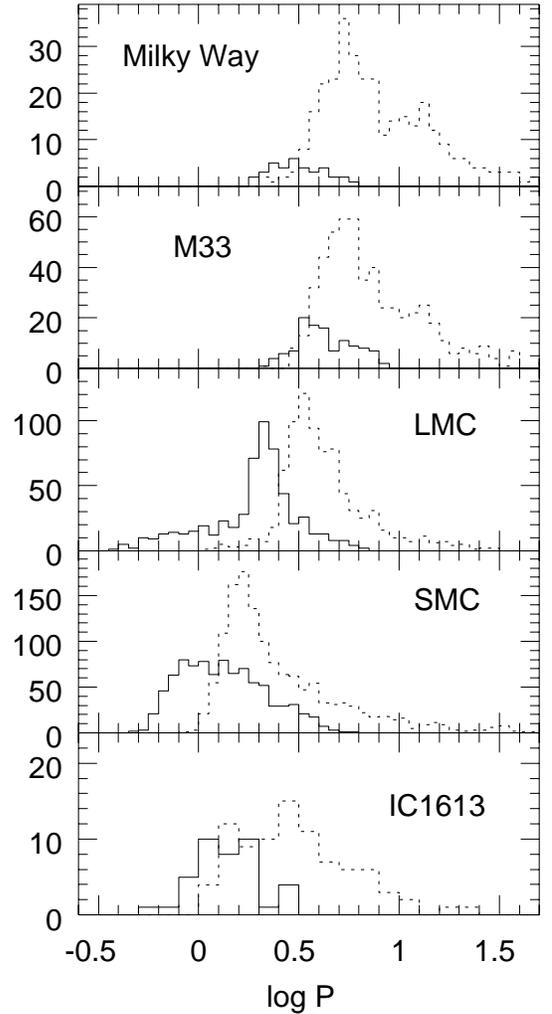}}
 \caption[ ]{Period distribution of Cepheids pulsating in $F$
(dotted line) and $1OT$ (continuous line) mode
}
 \label{isto1}
\end{figure}

\begin{figure}
\resizebox{\vsize}{!}{\includegraphics{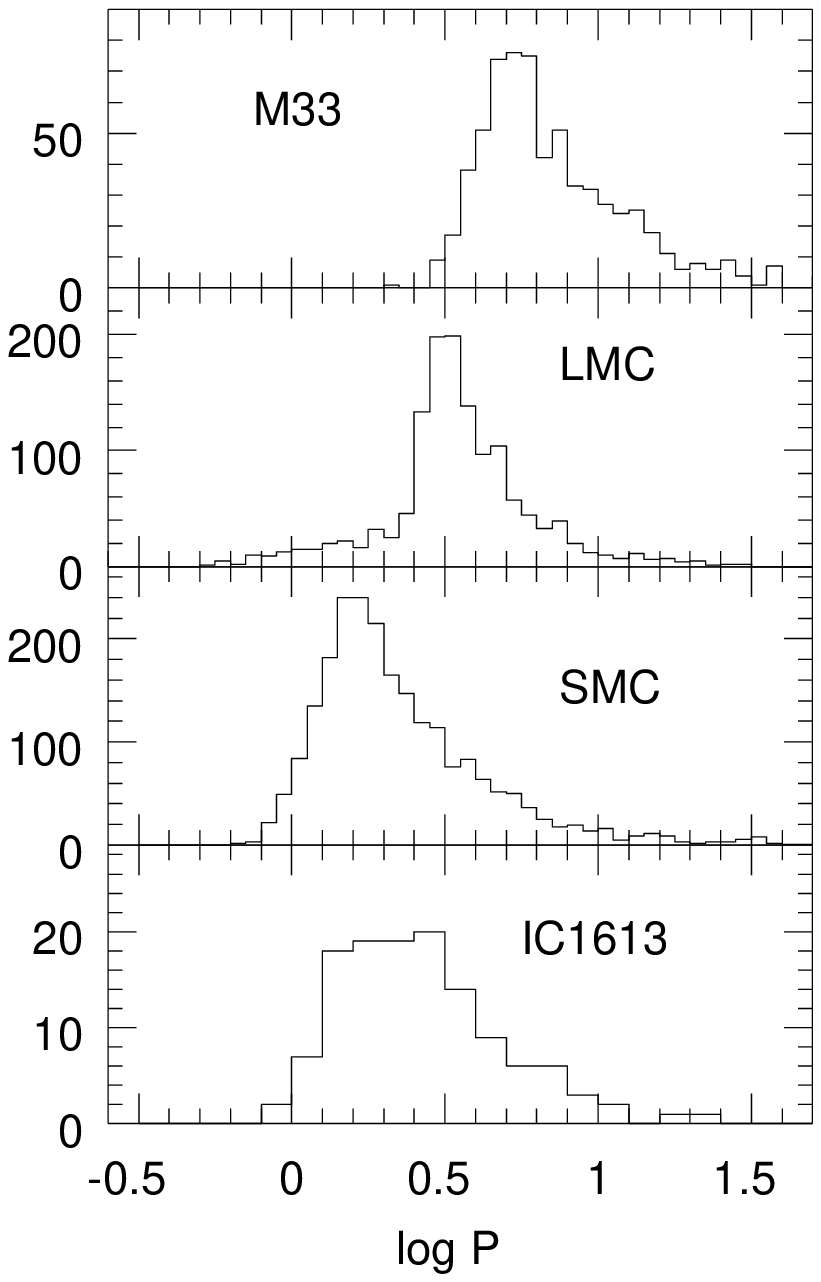}}
 \caption[ ]{Homogeneous period distribution for $F$ and $1OT$ mode
Cepheids; the $P$ of $1OT$ pulsators was multiplied by a factor $1/0.7$,
which is an average value of the ratio $P_0/P_1$ (see text)
}
 \label{isto2}
\end{figure}

The $P$ distribution is shown in Figs. \ref{isto}, \ref{isto1} and \ref{isto2} 
and is compared with those of other galaxies in the Local Group. In Fig. 
\ref{isto} the histogram of all the classical Cepheids is reported, while
in Fig. \ref{isto1} the stars are separated according to the pulsation mode;
finally, Fig. \ref{isto2} shows the homogeneous distribution when the $P$ of
$1OT$ pulsators is multiplied by an average period ratio $P_0/P_1 \sim 1/0.7$.
The adopted bin size is $\Delta{\log P}=0.05$; for IC 1613 in 
Figs. \ref{isto1} and \ref{isto2} it is $\Delta{\log P}=0.1$. Since we are
interested in pointing out the new results with respect to previous studies, 
the histograms are focused on intermediate and shorter $P$ Cepheids 
($P \la 50$ d); a comment on longer $P$ stars is however reported as a 
conclusive note in the Discussion.

{\em IC 1613}. The histogram shown in Fig. \ref{isto} includes all the known
209 Cepheids, while the sample in Figs. \ref{isto1} and \ref{isto2} is based
on our survey and it includes the stars found by Dolphin et al. (\cite{dol}). 
The samples should be considered almost complete for $P$ longer than about 
2 - 2.5 d ($\log{P} \sim 0.4$). Probably there are no $1OT$ Cepheids with 
$P \ga 3$ d.

{\em Milky Way}. For the histogram in Fig. \ref{isto} we considered 504 
stars in the Galactic Cepheid Database (Fernie et al. \cite{fer}). For 
Fig. \ref{isto1} we have used the results on the 348 stars with good light 
curves analyzed by the Brera-Merate group (e.g. Poretti \cite{por}, Antonello
\& Morelli \cite{am}, and references therein), which include 36
identified $1OT$ pulsators. We recall that it is possible to discriminate 
between $1OT$ mode and $F$ mode stars using light curve Fourier parameters 
only when the $P$ is shorter than about 5.5 d, since above this value the
light curves tend to be similar. Therefore no identified $1OT$ mode stars 
exist with longer $P$. On the other hand, the identified $1OT$ 
Cepheids whose distribution is shown in Fig. \ref{isto1} should constitute a 
fairly complete sample of these stars with $P \la 5.5$ d in the 
vicinity of the Sun.

{\em LMC and SMC}. The data for stars in LMC and SMC were taken from OGLE 
data archive (Udalski et al. \cite{uda2}, Udalski et al. \cite{uda3}, Udalski 
et al. \cite{uda4}, Soszynski et al \cite{sos}); they include also the double 
mode Cepheids (DMC). For SMC we have 1343 $F$ stars (24 of them are DMCs 
pulsating in $F$ and $1OT$ mode), and 890 $1OT$ stars (71 DMCs pulsating in
$1OT$ and $2OT$ mode). For the LMC we have 791 $F$ stars (20  DMCs 
pulsating in $F$ and $1OT$ mode), and 569 $1OT$ stars (61 DMCs pulsating in
$1OT$ and $2OT$ mode). Owing to OGLE observational constraints, the sample 
does not contain stars with $P$ longer than about 40 d; however this is not an 
issue for the problem we will discuss here. On the other hand, the sample
is fairly complete as regards the shorter $P$ stars, and is well 
representative of the Cepheid population in the surveyed fields of the
LMC and SMC. The 
distributions of LMC Cepheids show a long tail at shorter $P$, given by 
$1OT$ pulsators.

{\em M31}. After an analysis of DIRECT M31 data archive (Mochejska et al.
\cite{m31} and references therein), we preferred to adopt the previous 
results on 392 stars reported in the GCVS (Samus \cite{sam}). The reason is 
that DIRECT sample appears incomplete for $P$ shorter than about 6 d, where 
the old data distribution shows a prominent maximum; the difference is not due 
to possible different observed regions of the galaxy. As discussed by BIT, 
the sample produced by the photographic data should be considered 
representative for $\log P \la 0.7$.
We note in passing that there is not yet information available on the 
existence of $1OT$ pulsators in this galaxy.

%
%
{\em M33}. For M33 we adopted the results of the DIRECT survey published
by  Macri et al. (\cite{mac}), Mochejska et al. (\cite{moc1}) and
Mochejska et al. (\cite{moc2}); the Cepheid lists were checked for the 
presence of the same star in different lists, on the basis of stellar 
coordinatae. The sample contains 654 stars. Given the adopted 
observational techniques, the Cepheids with $P > 14$ d are partially 
underestimated. On the whole, keeping in mind this possible selection effect,
we think the sample is representative of the Cepheid population 
for $\log P \ga 0.5$.
Unfortunately the authors do not report the mode classification, even if 
uncertain; the $PL$ relation for M33 shows the very probable presence of $1OT$
mode pulsators with $P$ up to 7 - 8 d. We have adopted the separation line
reported in their Figure 11 for discriminating between $F$ and candidate
$1OT$ mode stars. Of course, the results of the analysis should be considered 
only as preliminary.

\section{Discussion and Conclusion}
We are trying to compare the frequency--period distribution in different 
galaxies even if the samples have different degrees of completeness. 
The possible selection effects that hinder attempts to achieve complete 
samples were discussed by BIT, and we can revise them, taking into account the
progresses made in the last twenty years. Variables with large amplitude 
variations are more likely to be discovered than those with small amplitude 
variations. Since the amplitude is related to the $P$, the discovery 
of stars in the $P$ ranges where the amplitude is larger is favored. The
photographic technique made it difficult to find Cepheids with amplitudes
less than about 0.5 mag; for the CCD technique this limit is much lower
(for example, this has allowed the detection of $1OT$ pulsators in galaxies 
beyond the Magellanic Clouds). The main selection effect is therefore the 
limiting magnitude of each survey, coupled with the detectable amplitude 
(which is related to the error of the measurements). This favors the selection
of Cepheids with longer $P$ since they are brighter, while the detectable 
Cepheids with shorter $P$ and for a given pulsation mode tend to be those with
larger amplitudes. In Sect. 3 we have reported for each galaxy the estimated 
shorter $P$ limit, and a reliable comparison between galaxies can be done 
above such a value. Finally, we mention the ever--present problem of 
discovering variables with $P$ close to 1 d; this difficulty is related to the 
constraint of observing at one-day intervals.

Many of the differences among the various galaxies in Figs. \ref{isto}, 
\ref{isto1} and \ref{isto2} can be interpreted in terms of different average 
metallicity. According to the data collected by van den Bergh (\cite{vdb}), 
the metallicity indicator values (for young population I stars) can be 
summarized as in Table 1. As is well known (see e.g. BIT), the location of 
the maximum of the frequency--period distribution increases with increasing 
metallicity.

As regards IC 1613, taking into account the incompleteness of the sample,
the distributions are in part similar to those of SMC. The 
maximum of the distribution in Fig. \ref{isto} would be probably located 
at $P \la $  1.6 d ($\log{P} \la 0.2$), and we should expect a larger 
number of stars with $P \la 2$ d yet to be discovered. The peak at 
$\log P = 0.4$ is intriguing, but we cannot give it much weight.
It is interesting to compare the LMC and SMC in Figs. \ref{isto} and 
\ref{isto2}; after roughly correcting for the different pulsation modes, 
the shape difference
of the distributions (Fig. \ref{isto}) decreases significantly (Fig. 
\ref{isto2}). On the other hand, the similarity of the shape of the
distributions of $1OT$ and $F$ pulsators in LMC (Fig. \ref{isto1}) appears 
rather unique among galaxies.
\begin{table}
\caption[]{Values of metallicity indicators of galaxies of the 
Local Group (young stars; van den Bergh, \cite{vdb})}
\begin{flushleft}
\begin{tabular}{lll}
\hline\noalign{\smallskip}
galaxy & $[Fe/H]$ & $12+\log{(O/H)}$ \\
\noalign{\smallskip}\hline\noalign{\smallskip}
M31       &       &  9.0    \\
Milky Way & +0.06 &  8.7    \\
M33       &       &  8.4    \\
LMC       & -0.30 &  8.37   \\
SMC       & -0.73 &  8.02   \\
IC 1613   & -1.3  &  7.86   \\
\noalign{\smallskip}
\hline
\end{tabular}
\end{flushleft}
\end{table}

In spite of the different metallicities, the location of
primary and secondary maxima of M31, Milky Way and M33 are practically the 
same. We can see in Fig. \ref{isto} that they are located at about 5 d 
($\log{P} \sim 0.7$) and 13 d ($\log{P} \sim 1.1$; more exactly, between 12.6 
and 14.1 d), respectively. The recent observations of M33 made by the DIRECT 
project give us the opportunity to update the study on the possible origin 
of the dip located at about 8 - 10 d ($\log{P} \sim 0.9 - 1.0 $). According 
to the interpretation of Buchler et al. (\cite{bu1}), in the Milky Way a 
deficiency of $F$ mode pulsators in this $P$ range is 
expected owing to the instability of the pulsation cycle; since such stars 
will actually pulsate in the $1OT$ mode, the deficiency must be compensated by 
an excess of $1OT$ pulsators in the $P$ range of about 5.6 - 7 d. 
As noted in the previous Section 3 (Milky Way), it is not possible to 
discriminate the pulsation mode for $P\ga 5.5$ d using light curve parameters
alone, therefore for the present we cannot say how many $1OT$
pulsators are actually located between $\log{P} \sim 0.75$ and 
$\sim 0.84$ in the Milky Way. In M33, according to the $PL$ diagram reported
by Mochejska et al. (\cite{moc2}), there are some probable $1OT$ stars 
located in such $P$ range (Fig. \ref{isto1}), but it is difficult to affirm 
that there exists an excess of $1OT$ stars;
on the other hand also the dip, which is located near 10 d, is not very 
pronounced. The exercise shown in Fig. \ref{isto2} indicates that indeed the 
dip tends to be filled by the $1OT$ pulsators, if we change their $P$ to that 
of the $F$ mode one, but in any case there is still a bimodal distribution, 
i.e. there is an excess of stars with $\log{P} \sim 1.1$ which must be 
explained. Finally, it is instructive to look at M31 even if there is no 
information on the $1OT$ pulsators. In this case, even assuming a certain 
number of $1OT$ pulsators with $P$ within 5.6 - 7 d which more or less fill 
the dip, they will never cancel the evident bimodal distribution.
It is important to remark that we do not exclude the effect proposed by
Buchler et al. (\cite{bu1}). We just say that it does not look
sufficient to explain the bimodal distribution. 

It is difficult to verify which of the two interpretations, 
BIT or Chiosi's (\cite{chi}) one, would be more acceptable. Roughly speaking, 
M33 data confirm the expectation of both that the secondary component 
depends on the metallicity of the galaxy. If we make the reasonable 
assumption that the samples are fairly complete for a sufficiently 
long $P$ (e.g. above 6 - 7 d), and we consider normalized distributions of
M33, Milky Way and M31, the data indicate that this component increases 
montonically with the metallicity. Then the question is which of the two 
hypotheses predicts this relationship. The answer requires a more detailed 
study. 

The clarification of some of these problems requires more accurate photometric
data of Cepheids in the Local Group galaxies. Moreover, the longer $P$ $1OT$
Cepheids in the Milky Way have yet to be identified; a useful technique for
such an identification could be the analysis of the radial velocity curves.
The preliminary results obtained by Moskalik \& Ogloza (\cite{mo1}) suggest 
indeed that the Fourier parameters of radial velocity curves and also the 
phase lag between light and radial velocity curves could be mode 
discriminators for $P \ga 5$ d. This is a necessary study before beginning
a reliable discussion of the Cepheids in the Milky Way. 
Unfortunately, we think there is a further dilemma: 
assuming the excess of $1OT$ pulsators does exist, should it be the expected 
signature of the pulsational stability or would it be just the 
bimodal distribution of the $1OT$ pulsators?

As a final remark we note that, with respect to BIT, there are few new
observational results on very long ($P > 100$ d) Cepheids. The HST Key Project 
on $H_0$ determination was a good opportunity for a statistics on a large 
sample of galaxies, but unfortunately the observations were limited to periods
shorter than 100 d. According to BIT, Cepheids with $\log P > 2.0$ are 
probably massive stars burning carbon in their cores, and those authors infer
that a whole galaxy such as M31 would have to be searched to find just one
Cepheid in the core--carbon--burning phase. However, these stars are present 
in dwarf irregular galaxies such as LMC, SMC, IC 1613 and, as recently
observed, also in NGC 6822 (Antonello et al. \cite{pa5}). BIT interpreted the
discrepancy as further support for their suggestion of a two--component
birthrate function, the second component giving 5 -- 50 times as many
massive stars as indicated by the one--component model. However, the nonlinear
pulsation characteristics could also play a role in this case.
Aikawa \& Antonello (\cite{aa}) found that the nonlinear pulsation cycle 
for $P > 100$ d is stable only for low metallicity values, therefore the
very long $P$ Cepheids should be preferably detected in metal--poor galaxies
such as dwarf irregulars. This would be a simple explanation of what is
actually observed.

\end{document}